\documentclass[fleqn,10pt]{wlscirep}
\usepackage[utf8]{inputenc}
\usepackage[T1]{fontenc}
\usepackage[english]{babel}
\usepackage{subcaption}
\usepackage{amsmath,amsfonts,amssymb}
\usepackage{graphicx}
\usepackage{xcolor}
\usepackage{soul}
\sethlcolor{yellow}
\usepackage[colorlinks=true, allcolors=blue]{hyperref}
\usepackage{appendix}

\title{Joint Encryption and Error Correction for Secure Quantum Communication}

\author[1,*]{Nitin Jha}
\author[1]{Abhishek Parakh}
\author[2]{Mahadevan Subramaniam}
\affil[1]{Kennesaw State University, GA, USA}
\affil[2]{University of Nebraska Omaha, NE, USA }

\affil[*]{Corresponding author: njha1@students.kennesaw.edu}

\begin{abstract}
Secure quantum networks are a bedrock requirement for developing a future quantum internet. However, quantum channels are susceptible to channel noise that introduce errors in the transmitted data. The traditional approach to providing error correction typically encapsulates the message in an error correction code after encryption. Such separate processes incur overhead that must be avoided when possible. We, consequently, provide a single integrated process that allows for encryption as well as error correction. This is a first attempt to do so for secure quantum communication and combines the Calderbank-Shor-Steane (CSS) code with the three-stage secure quantum communication protocol. Lastly, it allows for arbitrary qubits to be transmitted from sender to receiver making the proposed protocol general purpose.
\end{abstract}
\begin{document}

\flushbottom
\maketitle
%
%
\thispagestyle{empty}

\section{Introduction}
In recent decades, quantum information science has gained much interest and has produced several breakthroughs such as quantum key distribution protocols (QKD)\cite{bennett2014quantum}, algorithms for large-integer decomposition \cite{RSA1978}, and searching.\cite{Grover1996} At the same time, several companies are developing large-scale quantum computers. \cite{Proctor2022Measuring} NIST anticipates that useful quantum computers will be developed within a decade or so and is therefore working to replace existing public-key cryptographic algorithms with new quantum-resistant cryptographic algorithms. \cite{NISTStatusReport} While this will secure the networks during the classical-quantum transition phase, in the future, we will need to securely connect the scalable quantum computers using purely quantum networks and protocols that enable the transmission of qubits in full fidelity (without quantum-classical-quantum conversion). Quantum Secure Direct Communication (QSDC) allows this.\cite{qsdc} Most of these protocols, however, require long-term qubit storage or quantum entanglement that may not be possible for a long time on a quantum network. Furthermore, most QSDC protocols only allow messages representing bit strings to be transmitted rather than securely transmitting arbitrary qubit states.\cite{anirbanReviewQKD2017} Kak's three stage-protocol secure quantum communication protocol \cite{Kak2006-3Stage} supports QSDC and requires no classical communication as a part of the protocol itself. Furthermore, it allows for QSDC to be implemented using currently available technology without the need for quantum memory and is stable over long distances since it tolerates multi-photon bursts. \cite{burr2022evaluating, jhaSPIE2024}

In general, quantum networks are affected by several different types of noises reducing the overall efficiency and security of the physical channels. Attackers can masquerade within a noisy environment to illicit useful information without detection. Therefore, secure quantum communication networks\cite{Preskill1998ReliableQC} will require error detection and mitigation techniques to make them reliable.\cite{Chiaverini2004} Quantum error correction\cite{CSS, Steane1996MultiplePI} is thus one of the most important steps moving forward. It involves identifying and correcting errors by exploiting certain quantum mechanical principles in conjunction with classical error correction methods to form quantum error correction codes (QECC). 

There are several proposed error correction codes constructed from classical error correction codes, such as the Hamming code, the BCH (Bose, Chaudhuri, and Hocquenghem) codes and the Reed-Solomon codes, which are not efficient at current noise levels. \cite{bb84css} As a result, low density parity check (LDPC) codes were introduced as quantum error correction codes, which are more efficient than the previous codes. \cite{Xu2020} Apart from these, \textit{Cascade} method was extensively used for error correction, where the communicating parties use parity checks of smaller segments and use binary search to locate the errors for protocols such as BB84.\cite{Xu2020, bb84css} Calderbank, Shor, and Steane proposed a class of quantum error correction codes, called CSS codes, that combines concepts from linear error correction codes and some properties of QECCs but this study was limited and inefficient.\cite{CSS} 

While QECC has been extensively studied and integrated with quantum computing architecture and paradigm, its integration with secure quantum communication schemes has been limited. One such study by (Bala, et al. 2023)\cite{Bala2023} laid down the formulation for a method to extract the invariants ({i.e., parameters that are only dependent on the initial state and not their respective noisy evolution}) that can be used to transmit data in an uncorrupted manner as the noisy channels have no effects on such parameters. In addition to QECCs, (Hu et al., 2019)\cite{Hu2019} proposes a quantum key distribution method based on \textit{ quantum error-avoiding code} using group theory based on \textit{decoherence-free subspace} (DFS) which, in theory, has been proven to have quantum error correction properties. Parakh studied the effect of rotational noise on the three-stage protocol by integrating basic protection against bit-flip errors.\cite{parakh2016correcting,parakh2018using} Wei et al. \cite{bb84css} propose a modified BB84 protocol using QECC but does not extend to QSDC schemes and only suitable for classical messages being transmitted using quantum protocols. 

Pan et al., \cite{Pan2020}  {utilized forward error-correction (FEC) codes to mitigate transmission errors and enhance the robustness of the quantum communication channel. They incorporated low-density parity check (LDPC) codes to handle the error rates for a suitable secure communication over free-space.} Sun et al.,\cite{Sun2020}  { proposed a quantum-memory-free protocol that leverages classical FEC codes to improve communication reliability, which showed the significance of using classical codes for desgining more efficient quantum error correction codes. Their protocol design underscores the significance of integrating classical error correction methods within quantum communication frameworks to maintain data integrity and reduce the impact of quantum noise. Both studies highlight the indispensable role of classical FEC codes in overcoming the practical challenges associated with quantum communication systems, thereby bridging the gap between theoretical models and real-world applications. } 

 {Recent developments in Quantum Secure Direct Communication (QSDC) have made substantial progress both experimentally and theoretically.} Sheng et al.,\cite{sheng2022one}  {introduced a one-step QSDC protocol that streamlines the communication process by combining encoding and transmission into a single step, enhancing efficiency and security by reducing complexity and potential vulnerabilities.} Paparelle et al., \cite{paparelle2023practical}  {advanced practical implementations of QSDC using squeezed states, which exploit the unique properties of squeezed light to improve robustness and security against noise, making QSDC more viable in real-world noisy environments.} Zhang et al., \cite{zhang2017quantum}  {integrated quantum memory into QSDC systems, allowing for the storage and retrieval of quantum states, thereby enhancing the scalability and robustness of quantum communication networks.}

In this paper, we propose the first joint quantum encryption and error correction technique for quantum protocols by extending the three-stage QKD protocol proposed by Kak\cite{Kak2006-3Stage}. The three-stage protocol allows QSDC using commuting unitary operations, private to communicating parties Alice and Bob, similar to double-lock cryptography, thereby removing the need for a classical channel and the need for a conversion of quantum signals to classical signals. Consequently, arbitrary qubit states output from algorithms running on a quantum computer can be transmitted in full fidelity to another quantum computer, far away, for further processing. The three-stage protocol does not have an inherent loss in qubits, since it does not measure qubits in random bases. 

The secret unitary transformations in the three-stage protocol act as encryption functions on a quantum state \cite{kakProof2009} since there are infinite possibilities for such a secret transformation. On the other hand, this also calls for an increased need for quantum error correction to ensure the correctness of the information being transmitted. Furthermore, certain types of errors may make the original secret commuting unitary transformations non-commuting leading to erroneous decoding of the qubit by the receiver. Consequently, a joint encryption and error correction technique is presented by extending the three-stage quantum key distribution protocol based on CSS error correction codes. The proposed protocol corrects single-bit flip and single-phase flip error. 


\paragraph{Quantum Error Correction Codes:}
Quantum channels are typically implemented using high-fidelity optical fiber channels. Optical fiber channels are affected by several physical noises that decrease the overall quality of the transmission. \cite{Braunstein1998} This requires the development of error correction codes, which will take the noisy or error-induced state back to its original form. For a quantum state passing through a noisy channel, which effects the state by the operator $\mathcal{E}$, can be written as, 
\begin{equation}
    \rho \to \rho' \equiv \mathcal{E}(\rho) = \sum_j E_j \rho E_j^\dagger,
\end{equation}
where the Kraus operators, $E_j$, satisfies the condition $\sum_j E_jE_j^\dagger =\mathbb{I}$\cite{Bala2023}, where $\mathbb{I}$ is the identity matrix and each $E_j$ represents $j^{th}$ noise model. 


The idea behind the error correction code is to take a small subspace of the Hermitian space and use it as the coding basis.  This subspace is chosen in such a way that it will reasonably cause all predicted errors to shift the coding subspace into the orthogonal error subspace. When a bit or phase flip error occurs, the measurement may be used to verify the error message, the operation can be adjusted to fix the bit, and it is ensured that the operation won't have an impact on the coded state. A QECC is represented as $Q:[[n,k,d]]$, which is a subspace of size $2^k$ of the larger Hermitian Space of size $2^n$, where the quantities $n$, $k$, and $d$ are,
\begin{enumerate}
    \item $n$ is the number of the physical qubits used to store the information.
    \item $k$ is the number of logical qubits protected by this code
    \item $d$ is the distance of the code, i.e., if $d=3$ then the code protects against $1$ bit-flip or phase flip error occurred.
\end{enumerate}

The Calderbank-Shor-Steane (CSS) quantum error correction code is constructed using the classical linear code theory. Using this, we encode the quantum state of $k$ qubits into $n$ qubits (given $k<n$) such that any error in $t\leq \left[ (d-1)/2\right]$ qubits can be measured and thus corrected without altering the encoded state. In our discussion, $d$ represents the minimum distance of the error correction code $Q$. Based on the theorem mentioned in the paper by (Calderbank et al., 1995)\cite{CSS}, for any two linear binary classical codes $C_1 = [n_1, k_1, d_1]$ and $C_2 = [n_2, k_2, d_2]$ with $C_2^\perp \subseteq C_1$ and $n \leq k_1+k_2$, where $C_2^\perp$ is the dual code of $C_2$. It is possible to construct a quantum error correction code (QECC) as follows:
\begin{equation}
Q: [[n, k=k_1+k_2, \min\{d_1, d_2\}]].
\label{css_Q}
\end{equation}
The corresponding basis states can be represented as, 
\begin{equation}
    \left\{  |c_w\rangle = \frac{1}{2^{\frac{n-k_1}{2}}} \sum_{w \in C^\perp_1} |w+v\rangle,  w\in C_2/C_1^\perp \right\},
    \label{qubit_encoding_css}
\end{equation}
where $C_2/C_1$ defines all the cosets of $C_1$ in $C_2$.  We denote the generator and parity-check matrix of $C_i$ for $i=\{1,2\}$, as $G_i$ and $H_i$, respectively.

\section{Results}
\label{Sec:3-stage}
{In order to make the three-stage} quantum protocol robust to channel noise, we modify it using CSS. Assume that Alice is the sender, Bob is the receiver, and Eve is the eavesdropper. Eve can measure transmitted qubits as well as launch a photon number-splitting attack. Alice and Bob can choose any orthogonal basis for encoding their information.  {The protocol presented below is for the transmission of an encryption key.} In general, however, Alice may transmit a qubit that is produced as the output of an algorithm she ran and wants to transmit that qubit as is, securely, to Bob for further processing.

For simplicity, assume Alice encodes using horizontal/vertical ($Z$-basis). Thus, $|0\rangle$ and $|1\rangle$ represent bits $0$ and $1$, respectively. We use CSS($C_1$, $C_2$) codes, which correct up to $t$ errors.\cite{bb84css}

The proposed scheme to integrate CSS codes with the three-stage protocol is as follows, 
\begin{enumerate}
    \item Alice chooses $N$ qubits randomly and encodes each of them in a given basis, $Z$-basis in our example. This choice of basis is global knowledge, as proposed in the original three-stage protocol.

    Alice starts with a qubit string $q$ that consists of $N+\delta$ random classical bits  {encoded into qubit blocks (each qubit block consist of the code-words encoding, i.e., for example 7-qubit encoding as per Steane's code) of size $N$. } 
    \begin{equation}
        |\psi\rangle  = \bigotimes_{k=1}^{(N+\delta)}|\psi_{q_k}\rangle,
        \label{eq:psiinit}
    \end{equation}
    where $q_k$ is the $k^{th}$ qubit of the string $q$ mentioned above.  {$\delta\geq0$ can be adjusted by the user to reduce the effect of attenuation loss of qubits to ensure successful transmission of at least $N$ qubits from Alice to Bob. $\delta=0$ is used when there is no attenuation loss or the protocol is used for QSDC.}  Also, for eq(\ref{eq:psiinit}) the following shows the encoding, 
    \begin{equation}
        |\psi_{q_k=0}\rangle = |0\rangle_L \ \text{and}\   |\psi_{q_k=1}\rangle = |1\rangle_L,
        \label{encod}
    \end{equation}
    where $|0\rangle_L$ and $|1\rangle_L$ denote the encoded states. 
    \item After preparing the qubits, Alice applies a secret rotation to all her qubits by an angle $\theta$.Note that this value of rotation is known only to Alice and the eavesdropper or Bob has no way of determining this operation independently. We can write this operation as $U_A^{\otimes (N+\delta)}(\theta)$. This denotes a tensor product of applying identical rotation to each of the elements of the state $|\psi\rangle$. The state that Alice now transmitted to Bob is depicted by eq(\ref{stage1}), 
    \begin{equation}
        |\psi'\rangle = U_A^{\otimes (N+\delta)}(\theta)|\psi\rangle,
        \label{stage1}
    \end{equation}
    \item Bob receives the state, $\rho = [U_A^{\otimes (N+\delta)}(\theta)|\psi\rangle\langle\psi|(U_A^\dagger)^{\otimes (N+\delta)}(\theta)]$. Without performing a measurement, he also applies a secret rotation of angle $\phi$ to the received state and transmits it back to Alice. His operation can be denoted by $U_B^{\otimes (N+\delta)}(\phi)$. The updated state is depicted in eq(\ref{stage2}), 
    \begin{equation}
        |\psi''\rangle = U_B^{\otimes (N+\delta)}(\phi)U_A^{\otimes (N+\delta)}(\theta)|\psi\rangle,
        \label{stage2}
    \end{equation}
    Note that similar to Alice's transformation, Bob's transformation operator also applies the same secret rotation to all of the individual elements of the received state. After this, Bob transmits the updated state back to Alice. 
    \item Now, Alice would like to reverse her transformation applied on the original state by applying the unitary transformation $(U_A^\dagger)^{\otimes (N+\delta)}$. This transformation essentially removes the effect of her original transformation as, 
    \begin{equation}
        (U_A^\dagger)^{\otimes (N+\delta)}U_A^{\otimes (N+\delta)} = \mathbb{I}^{\otimes (N+\delta)},
        \label{eq:unitarymultiplicaiton}
    \end{equation}
    It can be easily shown that the unitary transformation, $U_A^{\otimes (N+\delta)}$ and $U_B^{\otimes (N+\delta)}$ are commuting in nature as they are shown to be in the general three-stage protocol.\cite{Kak2006-3Stage} Alice sends the updated state back to Bob. 
    \item Bob now applies the reverse of his unitary operation, i.e. $(U_B^\dagger)^{\otimes (N+\delta)}$ to the received state. He thus recovers the original state (he might receive the following state, $\{ \rho = \epsilon|\psi\rangle\langle\psi|\}$). Bob uses error correction scheme using ancilla qubit to identify and correct bit-flip or phase flip error. Once all the error syndromes are corrected, he goes on to measure the qubits in the globally declared basis of choice, i.e., $Z$-basis in this example. $\epsilon$ represents the operation affecting the state which can be a combination of the channel noises or a result of eavesdropping

      The benefit of three-stage protocol is the higher efficiency provided due to the use of the unitary matrices, which removes the need for two initial bases of encoding. This allows Alice and Bob to ideally retain $4n$ qubits of their result, as there would be no discarding due to mismatch in encoding basis. Assuming a loss of qubits to the environment, keeping appropriate value of $\delta$, Alice and Bob should retain about $4n$ qubits-- which is significant than the retained $2n$ qubits while using the BB84 protocol\cite{bennett2014quantum}.
    \item While the three-stage protocol was being conducted, Alice chooses one $v_k \in C_1$ at random. 
    
    Alice essentially chooses $4n$ bits from $C_1$ to form $v_k$, and randomly creates $4n$ qubits in the state $|0\rangle$ or $|1\rangle$ based on the chosen classical bits. 
    \item Alice and Bob selects part of the received qubits to conduct the sifting process for security threat identification.

    Once, Bob has measured the qubits (he should measure $4n$ qubits with high probability, after considering the loss of $\delta$ qubits to the environment, etc), Alice chooses a part of the qubits $\gamma$ (such that $\gamma \leq 2n$) for sifting process to identify any discrepancies in the system). They agree on a higher bound of error qubits $t$, and if the number of mismatches in the sifting process is greater than this bound ($t$), the round is terminated and the protocol starts from the beginning.

    \item Alice and Bob discard the qubits used for sifting as part of security amplification, so Alice is left with a $(4n-\gamma)$ qubit string of $x$ and Bob is left with $x+\epsilon$.
    \item Alice declares $x-v_k$ on the public channel, and Bob subtracts this from his bits, thus identifying the error syndrome and, hence, correcting it with the $C_1$ code to retrieve the form of $v_k$. Both Alice and Bob calculate the coset of $v+C_2$ in $C_1$ to get the key $\mathrm{k}$. 
\end{enumerate}

Once Alice and Bob gets an upper bound on the mutual information between Eve's data $E$ and the their respective data $A$ and $B$, we can do the following information reconciliation. Suppose that Bob receives $B = A+\epsilon $, where $\epsilon$ is the error received, either due to eavesdropping or channel noise. Since the protocol was executed properly that bounds the error as fewer than $t$. Thus, if Alice and Bob correct their state to the given code in $C_1$, and thus their result $A', B'\in C_1$ will thus be same, i.e., $W=A'=B'$.\cite{bb84css} This steps concludes the step for information reconciliation; however the mutual information of Eve about $W$ maybe very high and thus unacceptable with respect to network security. To reduce this error, Alice and Bob have to calculate the coset of $C_2$ in $C_1$, that is, $W+C_2$ in $C_1$. Thus, exploiting the Eve's lack of knowledge about $C_2$ and the error correcting nature of the code-space, the mutual information of Eve can be reduced, and thus privacy amplification is achieved.

\paragraph{Example:} 
    Consider that Alice wants to send a message to Bob using the modified three-stage secure quantum communication protocol. They start by defining $C_1$ and $C_2$ and use them to define and encode the state, $|\psi\rangle$ based on eq(\ref{eq:psiinit}) as follows.

Let Alice's secret message be $S_A = 10101$. She encodes this as qubits based on eq(\ref{eq:psiinit}) (and encoding in $Z-$basis) as, 
\begin{equation}
    |\Psi\rangle = |1_L\rangle\otimes|0_L\rangle\otimes|1_L\rangle\otimes|0_L\rangle\otimes|1_L\rangle,
    \label{eq:aliceexample}
\end{equation}
where $|0_L\rangle$ and $|1_L\rangle$ are states based on the construction of a chosen [[n,k,d]] code. For Steane $[[7,1,3]]$ code the qubits are encoded in the following basis, 
\begin{equation*}
    |0_L\rangle = \frac{1}{\sqrt{8}} (|0000000\rangle + |0001111\rangle + |0110011\rangle + |0111100\rangle + |1010101\rangle + |1011010\rangle + |1100110\rangle + |1101001\rangle),
\end{equation*}
and 
\begin{equation*}
    |1_L\rangle = \frac{1}{\sqrt{8}} (|1111111\rangle + |1110000\rangle + |1001100\rangle + |1000011\rangle + |0101010\rangle + |0100101\rangle + |0011001\rangle + |0010110\rangle).
\end{equation*}
Basically $|0_L\rangle$ is collection of all even parity terms (even numbers of $1$) and $|1_L\rangle$ is collection of all odd parity terms (odd numbers of $1$). Now, she wants to rotate each of the qubits state by a given angle $\theta$, known only to Alice. She does this preparing a unitary operation given by, 
\begin{equation}
    U_A^\otimes = U_A\otimes U_A\otimes U_A\otimes U_A\otimes U_A,
    \label{eq:u_a}
\end{equation}
and each $U_A$ can be written as, 
\begin{equation}
    U_A = R_A\otimes R_A \otimes R_A\otimes ... R_A,
\end{equation}
where $U_A$ is the $n^{th}$ fold tensor product based on a $[n,k,d]$ chosen code and $R_A$ is the rotation operator of the form,

\begin{equation}
    R_A = \begin{pmatrix}
    \cos(\theta) & -\sin(\theta) \\
    \sin(\theta) & \cos(\theta)
    \end{pmatrix},
\end{equation}
here $\theta$ is Alice's secret rotation. This mathematical formulation ensures that the entire qubit state is rotated by a given angle. Now, once applied the updated state can be written as, 
\begin{equation}
    |\Psi'\rangle = U_A^\otimes|\Psi\rangle.
    \label{stage-1}
\end{equation}
Now, Bob receives the state $\rho$, which represents the density matrix of the transmitted data by Alice. He applies his secret rotation by an angle $\phi$ and denoted by $U_B$ of the same form as Alice's secret rotation.


\noindent The resulting state can be written as, 
\begin{equation}
    |\Psi''\rangle = U_B^\otimes U_A^\otimes |\Psi\rangle.
    \label{eq:stage 2}
\end{equation}
 Now, for the third and last stage of transmission, Alice removes her rotation by applying the inverse of the original rotation operator, i.e., $(U^\dagger_A)^\otimes$. 
 
It's trivial to show that $U_A^\otimes$ and $U_B^\otimes$ are, in fact, commuting. Thus the updated state that Bob receives is represented as,
\begin{equation}
    |\Psi'''\rangle = U_B^\otimes|\Psi\rangle.
    \label{stage3}
\end{equation}
The transmission and all three stages have been concluded. Bob removes the rotation in the last stage. Now, Bob uses an ancillary qubit (in state $|0\rangle$) to identify the error syndrome without performing a measurement and collapsing the states. The stabilizers for the Steane code, derived from the classical [7,4,3] Hamming code, are as follows,
\begin{equation}
    \begin{array}{|c|ccccccc|}
     ine
\text{Stabilizers} & Q_1 & Q_2 & Q_3 & Q_4 & Q_5 & Q_6 & Q_7 \\
 ine
X_1X_3X_5X_7 & X & I & X & I & X & I & X \\

X_2X_3X_6X_7 & I & X & X & I & I & X & X \\

X_4X_5X_6X_7 & I & I & I & X & X & X & X \\
 ine
Z_1Z_3Z_5Z_7 & Z & I & Z & I & Z & I & Z \\

Z_2Z_3Z_6Z_7 & I & Z & Z & I & I & Z & Z \\

Z_4Z_5Z_6Z_7 & I & I & I & Z & Z & Z & Z \\
 ine
\end{array}
\label{eq:stabilizer}
\end{equation}
Based on eq(\ref{eq:stabilizer}), Bob can perform certain controlled gate operations to determine the error syndrome. Like he can perform $X$-gate operations on qubits 1, 3, and 7 based on stabilizer class one. If he gets all eigenvalues to be $+1$ after this operation and measurement on the ancillary qubit, then there's no error. However, the presence of $-1$ eigenvalue indicates the presence of error on the given qubit. Then he can perform the respective error correction methods, i.e., use $X-$gate to correct bit-flip errors and $Z-$ gate then $X-$gate to correct phase errors. Once error correction is concluded Bob measures the state in the $Z-$basis as discussed between Alice and Bob before the transmission stages. Now, Bob has measured a state $|\Phi\rangle$, which we an assume to be the following, 
\begin{equation}
    |\Phi\rangle = 10100.
\end{equation}
Let's say they use the last two bits for sifting process. We assume that for a longer bit string the QBER found from sifting is under the threshold value, Alice and Bob considers the protocol to be successful. Assuming the protocol is successfully executed, and Alice and Bob ends up with $x$ and $x+\epsilon$ respectively,
\begin{equation}
    x  = 101\ \text{and}\ (x+\epsilon) = 101.
    \label{eq:xx+e}
\end{equation}
Now, based on the discussed $C_1$ and $C_2$ code-space agreed upon by Alice and Bob, Alice randomly chooses $v_k \in C_1$. For this example, let's say $v_k$ is as follows, 
\begin{equation}
    v_k = 011.
    \label{eq:v_k}
\end{equation}
This is an abstract example, and we known that in reality this will be a code-word from $C_1$. Now, Alice computes $x-v_k$, 
\begin{equation}
    x\oplus v = 110.
    \label{eq:transmission}
\end{equation}
Bob can now determine the form of $v_k$ by subtracting the declared $x+v_k$ by Alice. Now for the final key calculation, and based on the form of $C_1$ and $C_2$, given $C_2^\perp \subseteq C_1$, we find a coset of $v+C_2 \in C_1$.

\section{Discussion}
\label{Sec:Security}

The steps of creating logical qubits and applying Alice's secret rotations can be combined into one step since all operations are unitary and commute. This is not possible with other protocols that either use shared entangled qubit pairs or convert back and forth between classical and quantum signal representations. The three-stage protocol, on the other hand, only uses single qubits traveling back and forth. The secret unitary transformations of the communicating parties act as encryption functions on quantum states allowing double-lock cryptography.

 {The relationship between the efficiency and complexity of the code depends on the amount of noise present in the channel. In the practical implementation of CSS codes, we need to find a relationship between the code rate (which defines the efficiency) and the distance (which defines the error-correction capability of a linear code). The code rate can be bound using the Gilbert-Varshamov (GV) bound for linear codes.}\cite{Gvbound}  {For a $[[n,k,d]]$ code, where $k=k_1-k_2$ and $d=\min(d_1,d_2)$, we can write the code rate as,}
\begin{equation}
    R = \frac{k_1-k_2}{n}.
    \label{code-rate}
\end{equation}
 {The GV gives the lower bound of $k$ for any linear code. The bound can be written as,}
\begin{equation}
    2^k \geq \frac{2^n}{\sum_{i=0}^{d-1} \binom{n}{i}}.
    \label{gv-bound}
\end{equation}
 {Similarly, for the dual code, we can write the GV bound as,}
\begin{equation}
    2^{n - k} \geq \frac{2^n}{\sum_{i=0}^{d_2^\perp-1} \binom{n}{i}}.
    \label{GV-dual}
\end{equation}
 {For linear codes $C_1$, $[[n,k_1,d_1]]$, and $C_2$, $[[n,k_2,d_2]]$, we can find bounds on $k_1$ and $k_2$ using} eq(\ref{gv-bound}).  {Taking $\log()$ on both sides and re-writing} eq(\ref{gv-bound}) and eq(\ref{GV-dual}),
\begin{equation}
k_1 \geq n - \log_2 \left( \sum_{i=0}^{d_1-1} \binom{n}{i} \right),
\label{k1}
\end{equation}
and
\begin{equation}
k_2 \geq n - \log_2 \left( \sum_{i=0}^{d_2-1} \binom{n}{i} \right).
\label{k2}
\end{equation}
 {Given $k=k_1-k_2$, we can we get the form of $k$ as,}
\begin{equation}
    k \geq \log_2 \left( \sum_{i=0}^{d_2-1} \binom{n}{i}\right) - \log_2 \left( \sum_{i=0}^{d_1-1} \binom{n}{i}\right).
    \label{k}
\end{equation}
 {Based on} eq(\ref{code-rate}),  {we can write the bound for code-rate as,}
\begin{equation}
    R \geq \frac{1}{n}\left (\log_2 \left( \frac{\sum_{i=0}^{d_2-1} \binom{n}{i}}{\sum_{i=0}^{d_1-1} \binom{n}{i} } \right) \right).
    \label{bound-coderate}
\end{equation}
Thus, eq(\ref{bound-coderate})  {provides a lower bound for a CSS code constructed using linear codes $C_1$ and $C_2$. This establishes a bound on the efficiency of the CSS code construction for a code with distance, $d$.}

However, adding redundancy to a security protocol has the potential to leak information. Security analysis conducted by Wei et al., 2019\cite{bb84css} can be used for the privacy amplification of our protocol and a similar method can be used to bound the mutual information between the legitimate parties and Eve by estimating the bound using collision entropy. The chance that two independent, identically distributed samples of $X$ are equal (a ``collision'') is measured by the collision entropy $H_c(X) = -\log_2 \left( \sum_{x \in \mathcal{X}} p(x)^2 \right)$. Based on theorem [3] by Wei et al., 2019\cite{bb84css}, we can define the bound on security using the collision entropy defined on a random variable as described below.

Given a random variable $X$ in the alphabet $\mathcal{X}$ with collision entropy $H_c(x)$ and probability distribution $p(x)$, and let $G$ be the random variable which is selected randomly from a generalized hash function class $\{0,1\}^m$ corresponding to $\mathcal{X}$, then
\begin{equation}
    H(G(X))|G)\geq H_c(G(X)|G)\geq m-2^{m-H_c(X)}.
    \label{collison-entropy}
\end{equation}
    
Theorem [3] can be used for privacy amplification for our protocol. The idea is to make Eve's uncertainty about the key, $\mathcal{S}$, as high as possible. We start with Alice and Bob publicly choosing a string from the alphabet, i.e., $g\in \mathcal{E}$ and applying it to $W$ ($W$ are the corrected strings that Alice and Bob have, i.e., $A'$ and $B'$ mentioned in Section [\ref{Sec:3-stage}]), achieving the new bit string $\mathcal{S}$. Given Eve's knowledge being $Z=z$, Eve's uncertainty about $W$ is given by collision entropy\cite{bb84css} and we can estimate the lower bound of some number $d$ using eq(\ref{collison-entropy}), then from Theorem [3] we can write, 
\begin{equation}
    H_c(\mathcal{S}|G, Z=z)\geq m-2^{m-d}.
    \label{h_c}
\end{equation}

From eq(\ref{h_c}), we can choose $m$ to be very small such that $2^{m-d}\to 0$, and thus $H \approx m$. This maximizes Eve's uncertainty about the key $\mathcal{S}$, thus maximizing the security, and in essence achieving privacy enhancement. During information reconciliation, Alice can compute several parity checks and send Bob a classical message $a$ (which has information about the subset specifications and the parity combinations chosen by Alice), so Bob can correct his message. This makes sure both Alice and Bob ends up with the same message, $W$. However, this gives Eve extra knowledge, given by $E=a$, which increases her collision entropy to $H_c(W|Z=z, E=a)$. This has a weak lower bound given by,

\begin{equation}
    H_c(W|Z=z, E=a)\geq H_c(W|Z=z)-H(E),
    \label{bound}
\end{equation}
where $H(E)$ is the \textit{Shannon Entropy}.\cite{bb84css} Although beyond the scope of this paper, this security analysis can be extended further to provide a more solid bound on the collision entropy, or a different approach can be used to estimate the leaked information.

 {Finally, it is to be noted that researchers}\cite{LDPC-CSS}  {have integrated low-density parity-check (LDPC) and CSS codes with the BB84 protocol. The integration was done by using arbitrary LDPC codes, and the authors concluded that the decoding performance was robust, and due to the use of arbitrary LDPC codes, the complexity was deemed lesser than other existing methods. The simulation results showed that this particular construction provides decodable codes in practical noise-levels and also bounds Eve’s mutual information about the keys. A similar construction can be done with a three-stage protocol, as it has been shown to be compatible with CSS through our work.}

\section{Conclusions}

This paper presents the first joint encryption and error correction scheme by integrating of the CSS quantum error correction code with the three-stage secure quantum communication protocol along with the security analysis for the proposed approach.  {The underlying three-stage protocol is more efficient than prepare and measure quantum protocols and equally (if not more) secure than the BB84 protocol} \cite{three-stageSecurityProof}.  {Consequently, the integration of three-stage protocol with CSS provides an option with higher security and efficiency than the integration of CSS with BB84}\cite{bb84css}. The three-stage protocol is an all-quantum protocol that allows for quantum secure direct communication and therefore of immense interest for future quantum networks. Future work can also include the study of different eavesdropping attacks, man-in-the-middle, and photon-number-splitting attacks.

\bibliography{main} 

\begin{thebibliography}{10}
\urlstyle{rm}
\expandafter\ifx\csname url\endcsname\relax
  \def\url#1{\texttt{#1}}\fi
\expandafter\ifx\csname urlprefix\endcsname\relax\def\urlprefix{URL }\fi
\expandafter\ifx\csname doiprefix\endcsname\relax\def\doiprefix{DOI: }\fi
\providecommand{\bibinfo}[2]{#2}
\providecommand{\eprint}[2][]{\url{#2}}

\bibitem{bennett2014quantum}
\bibinfo{author}{Bennett, C.~H.} \& \bibinfo{author}{Brassard, G.}
\newblock \bibinfo{journal}{\bibinfo{title}{Quantum cryptography: Public key distribution and coin tossing}}.
\newblock {\emph{\JournalTitle{Theoretical Computer Science}}} \textbf{\bibinfo{volume}{560}}, \bibinfo{pages}{7--11} (\bibinfo{year}{2014}).

\bibitem{RSA1978}
\bibinfo{author}{Rivest, R.}, \bibinfo{author}{Shamir, A.} \& \bibinfo{author}{Adleman, L.~M.}
\newblock \bibinfo{journal}{\bibinfo{title}{A method for obtaining digital signatures and public-key cryptosystems}}.
\newblock {\emph{\JournalTitle{Commun. ACM}}} \textbf{\bibinfo{volume}{21}}, \bibinfo{pages}{120--126} (\bibinfo{year}{1978}).

\bibitem{Grover1996}
\bibinfo{author}{Grover, L.~K.}
\newblock \bibinfo{title}{Fast quantum mechanical algorithm for database search}.
\newblock In \emph{\bibinfo{booktitle}{Proceedings of the Twenty-eighth Annual ACM Symposium on Theory of Computing}}, \doiprefix\url{10.1145/237814.237866} (\bibinfo{year}{1996}).

\bibitem{Proctor2022Measuring}
\bibinfo{author}{Proctor, T.}, \bibinfo{author}{Rudinger, K.}, \bibinfo{author}{Young, K.}, \bibinfo{author}{Nielsen, E.} \& \bibinfo{author}{Blume-Kohout, R.}
\newblock \bibinfo{journal}{\bibinfo{title}{Measuring the capabilities of quantum computers}}.
\newblock {\emph{\JournalTitle{Nature Physics}}} \textbf{\bibinfo{volume}{18}}, \bibinfo{pages}{75--79} (\bibinfo{year}{2022}).

\bibitem{NISTStatusReport}
\bibinfo{author}{Alagic, G.} \emph{et~al.}
\newblock \bibinfo{title}{Status report on the first round of the nist post-quantum cryptography standardization process}.
\newblock \bibinfo{type}{Tech. Rep.}, \bibinfo{institution}{National Institute of Standards and Technology} (\bibinfo{year}{2019}).
\newblock \bibinfo{note}{NIST Interagency/Internal Report (NISTIR) - 8240}.

\bibitem{qsdc}
\bibinfo{author}{Pan, D.} \emph{et~al.}
\newblock \bibinfo{journal}{\bibinfo{title}{The evolution of quantum secure direct communication: On the road to the qinternet}}.
\newblock {\emph{\JournalTitle{IEEE Communications Surveys \& Tutorials}}} \bibinfo{pages}{1--1}, \doiprefix\url{10.1109/COMST.2024.3367535} (\bibinfo{year}{2024}).

\bibitem{anirbanReviewQKD2017}
\bibinfo{author}{Shenoy-Hejamadi, A.}, \bibinfo{author}{Pathak, A.} \& \bibinfo{author}{Radhakrishna, S.}
\newblock \bibinfo{journal}{\bibinfo{title}{Quantum cryptography: Key distribution and beyond}}.
\newblock {\emph{\JournalTitle{Quanta}}} \textbf{\bibinfo{volume}{6}}, \bibinfo{pages}{1--47}, \doiprefix\url{10.12743/quanta.v6i1.57} (\bibinfo{year}{2017}).

\bibitem{Kak2006-3Stage}
\bibinfo{author}{Kak, S.}
\newblock \bibinfo{journal}{\bibinfo{title}{A three-stage quantum cryptography protocol}}.
\newblock {\emph{\JournalTitle{Foundations of Physics Letters}}} \textbf{\bibinfo{volume}{19}}, \bibinfo{pages}{293--296} (\bibinfo{year}{2006}).

\bibitem{burr2022evaluating}
\bibinfo{author}{Burr, J.}, \bibinfo{author}{Parakh, A.} \& \bibinfo{author}{Subramaniam, M.}
\newblock \bibinfo{title}{{Evaluating different topologies for multi-photon quantum key distribution}}.
\newblock In \bibinfo{editor}{Donkor, E.}, \bibinfo{editor}{Hayduk, M.}, \bibinfo{editor}{Frey, M.~R.}, \bibinfo{editor}{Jr., S. J.~L.} \& \bibinfo{editor}{Myers, J.~M.} (eds.) \emph{\bibinfo{booktitle}{Quantum Information Science, Sensing, and Computation XIV}}, vol. \bibinfo{volume}{12093}, \bibinfo{pages}{1209309}. \bibinfo{organization}{International Society for Optics and Photonics} (\bibinfo{publisher}{SPIE}, \bibinfo{address}{Orlando, Florida, United States}, \bibinfo{year}{2022}).

\bibitem{jhaSPIE2024}
\bibinfo{author}{Jha, N.}, \bibinfo{author}{Parakh, A.} \& \bibinfo{author}{Subramaniam, M.}
\newblock \bibinfo{title}{{Effect of noise and topologies on multi-photon quantum protocols}}.
\newblock In \bibinfo{editor}{Hemmer, P.~R.} \& \bibinfo{editor}{Migdall, A.~L.} (eds.) \emph{\bibinfo{booktitle}{Quantum Computing, Communication, and Simulation IV}}, vol. \bibinfo{volume}{12911}, \bibinfo{pages}{129110G}, \doiprefix\url{10.1117/12.3000586}. \bibinfo{organization}{International Society for Optics and Photonics} (\bibinfo{publisher}{SPIE}, \bibinfo{year}{2024}).

\bibitem{Preskill1998ReliableQC}
\bibinfo{author}{Preskill, J.}
\newblock \bibinfo{journal}{\bibinfo{title}{Reliable quantum computers}}.
\newblock {\emph{\JournalTitle{Proceedings of the Royal Society of London. Series A: Mathematical, Physical and Engineering Sciences}}} \textbf{\bibinfo{volume}{454}}, \bibinfo{pages}{385--410} (\bibinfo{year}{1998}).

\bibitem{Chiaverini2004}
\bibinfo{author}{Chiaverini, J.} \emph{et~al.}
\newblock \bibinfo{journal}{\bibinfo{title}{Realization of quantum error correction}}.
\newblock {\emph{\JournalTitle{Nature}}} \textbf{\bibinfo{volume}{432}}, \bibinfo{pages}{602--605} (\bibinfo{year}{2004}).

\bibitem{CSS}
\bibinfo{author}{Calderbank, A.~R.} \& \bibinfo{author}{Shor, P.~W.}
\newblock \bibinfo{journal}{\bibinfo{title}{Good quantum error-correcting codes exist}}.
\newblock {\emph{\JournalTitle{Physical Review A}}} \textbf{\bibinfo{volume}{54}}, \bibinfo{pages}{1098--1105} (\bibinfo{year}{1995}).

\bibitem{Steane1996MultiplePI}
\bibinfo{author}{Steane, A.}
\newblock \bibinfo{journal}{\bibinfo{title}{Multiple particle interference and quantum error correction}}.
\newblock {\emph{\JournalTitle{Proceedings of the Royal Society of London. Series A: Mathematical and Physical Sciences}}} \textbf{\bibinfo{volume}{452}}, \bibinfo{pages}{2551--2577} (\bibinfo{year}{1996}).

\bibitem{bb84css}
\bibinfo{author}{Jia, W.}, \bibinfo{author}{Feng, B.}, \bibinfo{author}{Yu, H.} \& \bibinfo{author}{Bian, Y.}
\newblock \bibinfo{title}{Quantum key distribution protocol based on css error correcting codes}.
\newblock In \emph{\bibinfo{booktitle}{ACM TURC Conference on Artificial Intelligence and Security}}, \bibinfo{pages}{8}, \doiprefix\url{10.1145/3321408.3326680} (\bibinfo{organization}{ACM}, \bibinfo{address}{Chengdu, China}, \bibinfo{year}{2019}).

\bibitem{Xu2020}
\bibinfo{author}{Xu, F.}, \bibinfo{author}{Ma, X.}, \bibinfo{author}{Zhang, Q.}, \bibinfo{author}{Lo, H.-K.} \& \bibinfo{author}{Pan, J.-W.}
\newblock \bibinfo{journal}{\bibinfo{title}{Secure quantum key distribution with realistic devices}}.
\newblock {\emph{\JournalTitle{Reviews of Modern Physics}}} \textbf{\bibinfo{volume}{92}}, \bibinfo{pages}{025002}, \doiprefix\url{10.1103/RevModPhys.92.025002} (\bibinfo{year}{2020}).

\bibitem{Bala2023}
\bibinfo{author}{Bala, R.}, \bibinfo{author}{Asthana, S.} \& \bibinfo{author}{Ravishankar, V.}
\newblock \bibinfo{journal}{\bibinfo{title}{Combating errors in quantum communication: an integrated approach}}.
\newblock {\emph{\JournalTitle{Scientific Reports}}} \textbf{\bibinfo{volume}{13}}, \bibinfo{pages}{2979}, \doiprefix\url{10.1038/s41598-023-30178-x} (\bibinfo{year}{2023}).

\bibitem{Hu2019}
\bibinfo{author}{Hu, Q.} \emph{et~al.}
\newblock \bibinfo{title}{A quantum key distribution scheme based on quantum error-avoiding code in decoherence-free subspace}.
\newblock In \emph{\bibinfo{booktitle}{Proceedings of the ACM Turing Celebration Conference-China}}, \doiprefix\url{https://doi.org/10.1145/3321408.3326681} (\bibinfo{year}{2019}).

\bibitem{parakh2016correcting}
\bibinfo{author}{Parakh, A.} \& \bibinfo{author}{van Brandwijk, J.}
\newblock \bibinfo{title}{Correcting rotational errors in three stage qkd}.
\newblock In \emph{\bibinfo{booktitle}{2016 23rd International Conference on Telecommunications (ICT)}}, \bibinfo{pages}{1--5}, \doiprefix\url{10.1109/ICT.2016.7500409} (\bibinfo{organization}{IEEE}, \bibinfo{address}{Thessaloniki, Greece}, \bibinfo{year}{2016}).

\bibitem{parakh2018using}
\bibinfo{author}{Parakh, A.}
\newblock \bibinfo{title}{Using fewer qubits to correct errors in the three-stage qkd protocol}.
\newblock In \emph{\bibinfo{booktitle}{Quantum Information Science and Technology IV}}, vol. \bibinfo{volume}{10803} (\bibinfo{organization}{SPIE}, \bibinfo{year}{2018}).

\bibitem{Pan2020}
\bibinfo{author}{Pan, D.}, \bibinfo{author}{Lin, Z.}, \bibinfo{author}{Wu, J.} \emph{et~al.}
\newblock \bibinfo{journal}{\bibinfo{title}{Experimental free-space quantum secure direct communication and its security analysis}}.
\newblock {\emph{\JournalTitle{Photonics Research}}} \textbf{\bibinfo{volume}{8}}, \bibinfo{pages}{1522--1531} (\bibinfo{year}{2020}).

\bibitem{Sun2020}
\bibinfo{author}{Sun, Z.}, \bibinfo{author}{Song, L.}, \bibinfo{author}{Huang, Q.} \emph{et~al.}
\newblock \bibinfo{journal}{\bibinfo{title}{Toward practical quantum secure direct communication: A quantum-memory-free protocol and code design}}.
\newblock {\emph{\JournalTitle{IEEE Transactions on Communications}}} \textbf{\bibinfo{volume}{68}}, \bibinfo{pages}{5778--5792} (\bibinfo{year}{2020}).

\bibitem{sheng2022one}
\bibinfo{author}{Sheng, Y.~B.}, \bibinfo{author}{Zhou, L.} \& \bibinfo{author}{Long, G.~L.}
\newblock \bibinfo{journal}{\bibinfo{title}{One-step quantum secure direct communication}}.
\newblock {\emph{\JournalTitle{Science Bulletin}}} \textbf{\bibinfo{volume}{67}}, \bibinfo{pages}{367--374} (\bibinfo{year}{2022}).

\bibitem{paparelle2023practical}
\bibinfo{author}{Paparelle, I.}, \bibinfo{author}{Mousavi, F.}, \bibinfo{author}{Scazza, F.} \emph{et~al.}
\newblock \bibinfo{journal}{\bibinfo{title}{Practical quantum secure direct communication with squeezed states}}.
\newblock {\emph{\JournalTitle{arXiv preprint arXiv:2306.14322}}}  (\bibinfo{year}{2023}).

\bibitem{zhang2017quantum}
\bibinfo{author}{Zhang, W.}, \bibinfo{author}{Ding, D.~S.}, \bibinfo{author}{Sheng, Y.~B.} \emph{et~al.}
\newblock \bibinfo{journal}{\bibinfo{title}{Quantum secure direct communication with quantum memory}}.
\newblock {\emph{\JournalTitle{Physical Review Letters}}} \textbf{\bibinfo{volume}{118}}, \bibinfo{pages}{220501} (\bibinfo{year}{2017}).

\bibitem{kakProof2009}
\bibinfo{author}{Chen, Y.}, \bibinfo{author}{Verma, P.~K.} \& \bibinfo{author}{Kak, S.}
\newblock \bibinfo{journal}{\bibinfo{title}{Embedded security framework for integrated classical and quantum cryptography services in optical burst switching networks}}.
\newblock {\emph{\JournalTitle{Security and Communication Networks}}} \textbf{\bibinfo{volume}{2}}, \bibinfo{pages}{546--554}, \doiprefix\url{https://doi.org/10.1002/sec.98} (\bibinfo{year}{2009}).
\newblock \eprint{https://onlinelibrary.wiley.com/doi/pdf/10.1002/sec.98}.

\bibitem{Braunstein1998}
\bibinfo{author}{Braunstein, S.~L.}
\newblock \bibinfo{journal}{\bibinfo{title}{Quantum error correction for communication with linear optics}}.
\newblock {\emph{\JournalTitle{Nature}}} \textbf{\bibinfo{volume}{394}}, \bibinfo{pages}{47--49}, \doiprefix\url{10.1038/27803} (\bibinfo{year}{1998}).

\bibitem{Gvbound}
\bibinfo{author}{Vu, V.} \& \bibinfo{author}{Wu, L.}
\newblock \bibinfo{journal}{\bibinfo{title}{Improving the gilbert-varshamov bound for q-ary codes}}.
\newblock {\emph{\JournalTitle{IEEE Transactions on Information Theory}}} \textbf{\bibinfo{volume}{51}}, \bibinfo{pages}{3200--3208} (\bibinfo{year}{2005}).

\bibitem{LDPC-CSS}
\bibinfo{author}{Ohata, M.} \& \bibinfo{author}{Matsuura, K.}
\newblock \bibinfo{title}{Constructing {{CSS}} codes with {{LDPC}} codes for the bb84 quantum key distribution protocol} (\bibinfo{year}{2007}).
\newblock \eprint{quant-ph/0702184v3}.

\bibitem{three-stageSecurityProof}
\bibinfo{author}{Chan, K.~W.}, \bibinfo{author}{El~Fifai, M.}, \bibinfo{author}{Verma, P.}, \bibinfo{author}{Kak, S.} \& \bibinfo{author}{Chen, Y.}
\newblock \bibinfo{journal}{\bibinfo{title}{Security analysis of the multi-photon three-stage quantum key distribution}}.
\newblock {\emph{\JournalTitle{International Journal on Cryptography and Information Security}}} \textbf{\bibinfo{volume}{5}}, \bibinfo{pages}{01--13}, \doiprefix\url{10.5121/ijcis.2015.5401} (\bibinfo{year}{2015}).

\end{thebibliography}

\section*{Authors Contributions}
N.J. is the primary author of the manuscript and received intellectual inputs from A.P. and M.S. to guide the research. All authors edited the manuscript.

\section*{Data availability statement}
All data generated or analysed during this study are included in this published article [and its supplementary information files].

\section*{Additional Information (Competing Interests)}
None

\end{document}